\PassOptionsToPackage{unicode}{hyperref}
\PassOptionsToPackage{hyphens}{url}
\documentclass[
  10pt,
]{article}
\usepackage{amsmath,amssymb}
\usepackage{iftex}
\ifPDFTeX
  \usepackage[T1]{fontenc}
  \usepackage[utf8]{inputenc}
  \usepackage{textcomp} 
\else 
  \usepackage{unicode-math} 
  \defaultfontfeatures{Scale=MatchLowercase}
  \defaultfontfeatures[\rmfamily]{Ligatures=TeX,Scale=1}
\fi
\usepackage{lmodern}
\ifPDFTeX\else
\fi
\IfFileExists{upquote.sty}{\usepackage{upquote}}{}
\IfFileExists{microtype.sty}{
  \usepackage[]{microtype}
  \UseMicrotypeSet[protrusion]{basicmath} 
}{}
\makeatletter
\@ifundefined{KOMAClassName}{
  \IfFileExists{parskip.sty}{%
    \usepackage{parskip}
  }{
    \setlength{\parindent}{0pt}
    \setlength{\parskip}{6pt plus 2pt minus 1pt}}
}{
  \KOMAoptions{parskip=half}}
\makeatother
\usepackage{xcolor}
\usepackage[margin=1in]{geometry}
\usepackage{color}
\usepackage{fancyvrb}

\DefineVerbatimEnvironment{Highlighting}{Verbatim}{commandchars=\\\{\}}
\newenvironment{Shaded}{}{}

\newcommand{\BuiltInTok}[1]{\textcolor[rgb]{0.00,0.50,0.00}{#1}}

\newcommand{\NormalTok}[1]{#1}
\newcommand{\OperatorTok}[1]{\textcolor[rgb]{0.40,0.40,0.40}{#1}}

\newcommand{\SpecialCharTok}[1]{\textcolor[rgb]{0.25,0.44,0.63}{#1}}
\newcommand{\SpecialStringTok}[1]{\textcolor[rgb]{0.73,0.40,0.53}{#1}}
\newcommand{\StringTok}[1]{\textcolor[rgb]{0.25,0.44,0.63}{#1}}
\newcommand{\VariableTok}[1]{\textcolor[rgb]{0.10,0.09,0.49}{#1}}

\usepackage{longtable,booktabs,array}
\usepackage{calc} 
\usepackage{etoolbox}
\makeatletter
\patchcmd\longtable{\par}{\if@noskipsec\mbox{}\fi\par}{}{}
\makeatother
\IfFileExists{footnotehyper.sty}{\usepackage{footnotehyper}}{\usepackage{footnote}}
\makesavenoteenv{longtable}
\usepackage{graphicx}
\makeatletter
\def\maxwidth{\ifdim\Gin@nat@width>\linewidth\linewidth\else\Gin@nat@width\fi}
\def\maxheight{\ifdim\Gin@nat@height>\textheight\textheight\else\Gin@nat@height\fi}
\makeatother
\setkeys{Gin}{width=\maxwidth,height=\maxheight,keepaspectratio}
\makeatletter
\def\fps@figure{htbp}
\makeatother
\providecommand{\tightlist}{%
  \setlength{\itemsep}{0pt}\setlength{\parskip}{0pt}}
\usepackage{microtype}
\usepackage{fvextra}
\fvset{breaklines=true,breakanywhere=true}
\usepackage{etoolbox}
\AtBeginEnvironment{longtable}{\footnotesize\setlength{\tabcolsep}{4pt}}
\AtBeginEnvironment{Highlighting}{\footnotesize}
\DefineVerbatimEnvironment{verbatim}{Verbatim}{breaklines=true,breakanywhere=true,fontsize=\footnotesize}
\ifLuaTeX
  \usepackage{selnolig}  
\fi
\IfFileExists{bookmark.sty}{\usepackage{bookmark}}{\usepackage{hyperref}}
\IfFileExists{xurl.sty}{\usepackage{xurl}}{} 
\hypersetup{
  hidelinks,
  pdfcreator={LaTeX via pandoc}}

\author{}
\date{}

\begin{document}

\hypertarget{a-challenge-nonce-freshness-gap-in-project-veraisons-tpm-reference-schemes-found-by-appraising-application-layer-action-evidence-end-to-end}{%
\section{A Challenge-Nonce Freshness Gap in Project Veraison's TPM
Reference Schemes, Found by Appraising Application-Layer Action Evidence
End-to-End}\label{a-challenge-nonce-freshness-gap-in-project-veraisons-tpm-reference-schemes-found-by-appraising-application-layer-action-evidence-end-to-end}}

\textbf{Anton Sokolov} Tyche Institute, Tallinn, Estonia Researcher ·
ORCID 0000-0003-2452-7096 · anton.sokolov@tyche.institute

\begin{center}\rule{0.5\linewidth}{0.5pt}\end{center}

\hypertarget{abstract}{%
\subsection{Abstract}\label{abstract}}

When an automated agent (an AI agent, say) takes a consequential action,
the record it leaves behind is produced by the very software stack whose
integrity is in question. A signed log proves which key wrote the
record, not what the runtime was. Prior work proposed treating that
record (an \emph{Action Evidence Package}, AEP: a signed append-only
record of an action, its authorising principal, and its outcome) as
application-layer Evidence under the IETF Remote ATtestation procedureS
(RATS) architecture (RFC 9334), binding the outcome into a
hardware-rooted TPM quote so that swapping it invalidates the quote. But
that prior work appraised the result only against a minimal Verifier
stand-in. This article closes that gap.

We drive an AEP quote, produced on an emulated software TPM, end-to-end
through a conformant Project Veraison RATS Verifier. We generate an EC
P-256 attestation key, measure the AEP outcome digest into a PCR, pack a
genuine quote into Veraison's \texttt{tpm-enacttrust} format, provision
a Concise Reference Integrity Manifest (trust anchor plus golden
reference value), and obtain a signed EAT Attestation Result (EAR). Good
evidence yields \texttt{affirming}; an outcome-swap or a one-byte
signature tamper yields \texttt{contraindicated}.

Along the way we uncover, responsibly disclose, and fix a
security-relevant finding: the reference scheme does not enforce
challenge-nonce freshness, so a replayed quote still appraises as
\texttt{affirming}. We give an exact, upstreamable two-part fix and
validate it end-to-end. With the fix active, the same valid quote that
is \texttt{affirming} in its own session flips to
\texttt{contraindicated} when replayed to a fresh one. The pipeline is
fully reproducible; the Attester remains an emulated \texttt{swtpm}, not
a hardware guarantee.

\textbf{Keywords:} remote attestation; RATS; Trusted Platform Module;
automated-agent accountability; evidence; Attestation Result; AR4SI;
EAR; CoRIM; Project Veraison; replay freshness; responsible disclosure

\begin{center}\rule{0.5\linewidth}{0.5pt}\end{center}

\hypertarget{introduction}{%
\subsection{1. Introduction}\label{introduction}}

\hypertarget{the-accountability-gap-for-automated-agent-actions}{%
\subsubsection{1.1 The accountability gap for automated-agent
actions}\label{the-accountability-gap-for-automated-agent-actions}}

Software agents now invoke tools, move money, file documents, and
reconfigure systems on behalf of human principals. When something goes
wrong, an after-the-fact review wants a faithful account: what the agent
did, who authorised it, and what the outcome was. The natural answer is
to have the agent emit a signed, append-only record of its actions. But
that answer has a structural defect that becomes acute exactly when it
matters most. The record is produced by the same software stack whose
integrity is in doubt. A later reviewer finds a faithful, signed
account, yet swapping the model for a cheaper one, or disabling a safety
filter, leaves the AEP well-formed, correctly signed, and wrong. The
witness is also the suspect {[}9{]}.

A digital signature answers the question \emph{which key wrote this?} It
does not answer \emph{what was the state of the machine that wrote it?}
Conflating those two questions is the heart of the accountability gap.
An attacker who controls the runtime can produce records that pass every
signature check and are nonetheless false.

The IETF Remote ATtestation procedureS (RATS) architecture, RFC 9334,
exists to separate these concerns. It splits apart the party that
produces measurements of its own state (the \textbf{Attester}), the
party that decides whether to believe those measurements (the
\textbf{Verifier}), and the party that acts on the resulting verdict
(the \textbf{Relying Party}). It adds an \textbf{Endorser} that vouches
for the Attester and a \textbf{Reference-Value Provider} that supplies
known-good measurements. The Verifier that judges the evidence is thus a
different party from the Attester that holds the secret being measured.
That separation is why RATS, rather than another signing scheme, is the
right substrate for this problem {[}9{]}.

The construct at the centre of this line of work is the \textbf{Action
Evidence Package (AEP)}: a signed, append-only record
\texttt{\{action,\ authorising\ principal,\ outcome\}} (plus inputs and
tool calls, hash-chained for tamper-evidence). The AEP is treated as
\emph{application-layer RATS Evidence} and bound to the platform's
hardware-rooted Evidence through \textbf{output-binding}. The outcome
digest is measured into a Platform Configuration Register that the quote
signs over, so the governance \emph{outcome}, and not merely a model
identity, is covered by the quote's signature, while the quote's
qualifying data carries the freshness nonce. (We use \textbf{AEP =
Action Evidence Package} throughout; this is unrelated to Sato's ``Agent
Execution Protocol,'' a distinct construct that must not be conflated.)
The composition pattern (AEP as application-layer Evidence appraised
alongside platform Evidence, with a two-axis verdict mapped to
standardised trustworthiness vocabulary) is set out in an
individual-submission IETF Internet-Draft (not WG-adopted; §3).

\hypertarget{what-was-missing-a-conformant-verifier}{%
\subsubsection{1.2 What was missing: a conformant
Verifier}\label{what-was-missing-a-conformant-verifier}}

The prior demonstration of output-binding had one deliberate,
clearly-marked limitation: it never reached a conformant Verifier. The
appraiser was an explicit stand-in for a Verifier such as Veraison. It
checked the signature, the freshness/qualifying data
(\texttt{tpm2\_checkquote}), and the quoted PCRs against a recorded good
state. What it did not do was appraise endorsement chains, ingest CoRIM
(Concise Reference Integrity Manifest) reference values, or produce
AR4SI/EAR results, and the exact Veraison service and interface names
awaited a real wiring-up. The prior work named that step as future work
{[}9{]}.

This article is that wiring-up. We replace the stand-in with
\textbf{Project Veraison} {[}7{]}, an Apache-licensed,
community-maintained reference implementation of a RATS Verifier, and
run our own \texttt{swtpm} AEP quote through its full provisioning and
challenge-response machinery to a signed, standards-conformant
Attestation Result. Doing so converts a \emph{provisional} verdict
mapping into one \emph{validated against a real EAR}, the validation the
Internet-Draft asks for (§3).

\hypertarget{contributions}{%
\subsubsection{1.3 Contributions}\label{contributions}}

This is a systems/security implementation-and-evaluation article. The
core result, in one sentence: a deployed, standards-conformant RATS
reference scheme silently omits challenge-nonce freshness while echoing
a nonce that masks the omission, and a one-line scheme patch plus a Rego
policy makes a byte-identical replayed quote flip from
\texttt{affirming} to \texttt{contraindicated}. We surface and validate
that finding by carrying automated-agent action evidence end-to-end
through a conformant Project Veraison Verifier. Foregrounding the
finding, the contributions are:

\begin{itemize}
\item
  \textbf{A challenge-nonce freshness gap in deployed RATS reference
  schemes, with a validated two-part fix (the headline finding).}
  Veraison's reference \texttt{tpm-enacttrust} scheme appraises
  signature and PCR-digest-versus-reference only and does \textbf{not}
  enforce challenge-nonce freshness, so a replayed or stale quote still
  returns \texttt{affirming}. We give a minimal, idiomatic fix (a
  one-line scheme patch that surfaces the attester-bound qualifying data
  as a claim, plus a Rego appraisal policy that compares it to the
  session nonce) and validate it end-to-end: with the fix active, the
  same valid quote that is \texttt{affirming} in its own session
  appraises as \texttt{contraindicated} when replayed to a fresh one. A
  source audit of the wider Veraison tree shows the omission spans
  \textbf{both} of its TPM-based reference schemes
  (\texttt{tpm-enacttrust}, which we validate end-to-end, and
  \texttt{parsec-tpm}, by source audit), while every
  platform-attestation scheme audited (\texttt{psa-iot},
  \texttt{arm-cca}, \texttt{sevsnp}) enforces freshness in its
  \texttt{AppraiseClaims}, so the gap is a scheme-family pattern and the
  fix is idiomatic. We disclose this responsibly and frame it as
  upstreamable (§6).
\item
  \textbf{The conformant-Verifier apparatus that surfaced it.} We
  provision and drive a real Project Veraison RATS Verifier, built and
  run in Docker on a single host, to appraise our own \texttt{swtpm} AEP
  quote, obtaining a signed EAR
  (\texttt{tag:github.com,2023:veraison/ear}, ES256): trust-anchor and
  reference-value provisioning via CoRIM, challenge-response, and
  AR4SI/EAR result production. Carrying an application-layer
  action-evidence construct through a conformant Verifier's full path,
  rather than the minimal stand-in of prior work, is what exposed the
  freshness gap and let us validate the verdict mapping against a real
  EAR (§4, §5).
\item
  \textbf{A verdict mapping confirmed at the status level.} Against real
  \texttt{swtpm} AEP evidence and a real Veraison appraisal, good
  evidence maps to \texttt{affirming}; an outcome swap whose PCR
  composite diverges from the golden reference maps to
  \texttt{contraindicated}; a signature tamper maps to
  \texttt{contraindicated}. We independently re-verify the affirming
  case (PCR digest equals the provisioned golden; signature verifies
  against the attestation-key public key). This confirms the prior
  provisional mapping at the submodule-status level for two of the three
  platform-axis terms (\emph{Attested} and \emph{Contested}) directly,
  and the third (\emph{Expired}/freshness) once the freshness fix of §6
  is active. One per-claim nuance, an AR4SI tier in the \emph{warning}
  band reported under a \texttt{contraindicated} submodule status, is
  surfaced as an open item rather than resolved (§5.1).
\item
  \textbf{A reproducible artifact.} A single driver script reproduces
  the whole pipeline (swtpm EC AK → measure → provision → quote → submit
  → EAR), and decoded EARs, submitted tokens, and an independent
  re-verifier are released, together with the provisioning gotchas that
  cost real time (§4, §7).
\end{itemize}

\hypertarget{scope-and-limitations-stated-up-front}{%
\subsubsection{1.4 Scope and limitations, stated up
front}\label{scope-and-limitations-stated-up-front}}

Two scope markers govern every claim in this article and are repeated
wherever they apply. First, the Attester is an \textbf{emulated software
TPM (\texttt{swtpm}), not a hardware guarantee}; a real discrete
STMicroelectronics dTPM proof-of-concept is staged for the IETF 126
Hackathon (Vienna, 18--19 July 2026), and we show in §9.1 that the
security finding does not depend on the emulated root. Second, the IETF
Internet-Draft on which the composition pattern rests is an
\textbf{individual submission, not WG-adopted}, and the IEEE pieces on
adjacent surfaces are \textbf{under review}. We carry these markers
verbatim wherever the claims they bound appear, and revisit them in the
limitations of §9.1.

\begin{center}\rule{0.5\linewidth}{0.5pt}\end{center}

\hypertarget{background}{%
\subsection{2. Background}\label{background}}

This section states the vocabulary precisely, expanding each
abbreviation on first use. Definitions of the AEP family draw on the
companion surfaces, cited where reused (§8).

\hypertarget{rats-roles-rfc-9334}{%
\subsubsection{2.1 RATS roles (RFC 9334)}\label{rats-roles-rfc-9334}}

The \textbf{RATS (Remote ATtestation procedureS)} architecture, RFC 9334
{[}1{]}, separates evidence production from evidence judgement:

\begin{itemize}
\tightlist
\item
  \textbf{Attester} --- produces \textbf{Evidence} about its own state
  (e.g., a TPM quote over measurement registers, signed by a
  hardware-rooted key).
\item
  \textbf{Verifier} --- appraises Evidence against \textbf{Reference
  Values} and \textbf{Endorsements} and emits \textbf{Attestation
  Results}.
\item
  \textbf{Relying Party} --- consumes the Results and makes a trust
  decision.
\item
  \textbf{Endorser} --- a vendor or authority that vouches for the
  Attester (e.g., that an attestation key belongs to a genuine TPM);
  supplies \textbf{Endorsements}.
\item
  \textbf{Reference-Value Provider} --- supplies known-good measurements
  (\textbf{Reference Values}) against which the Verifier compares
  Evidence.
\end{itemize}

The point of the architecture is that the Verifier's appraisal is
independent of the Attester's self-assertion: the Attester cannot, by
signing, make a measurement true.

\hypertarget{tpm-pcr-quote-ak-ek-qualifying-data}{%
\subsubsection{2.2 TPM, PCR, quote, AK, EK, qualifying
data}\label{tpm-pcr-quote-ak-ek-qualifying-data}}

A \textbf{TPM (Trusted Platform Module)} is a hardware (or emulated)
security chip that holds keys and accumulates measurements, specified by
the TCG TPM 2.0 Library Specification {[}13{]}. A \textbf{PCR (Platform
Configuration Register)} is a TPM register that can only be
\emph{extended} (\texttt{PCR\_new\ =\ H(PCR\_old\ ‖\ measurement)}), so
its value reflects the exact ordered sequence of measured components.

A \textbf{quote} is a TPM operation that signs, with a TPM-held key, a
digest over a selected set of PCRs together with caller-supplied
\textbf{qualifying data}, yielding two structures: \texttt{TPMS\_ATTEST}
(the signed-over attestation data, which includes the PCR digest and the
qualifying data in its \texttt{ExtraData} field) and
\texttt{TPMT\_SIGNATURE} (the signature). An \textbf{AK (Attestation
Key)} is a restricted TPM signing key used to sign quotes; in this work
a fresh \textbf{EC P-256 ECDSA} AK in \texttt{swtpm}. An \textbf{EK
(Endorsement Key)} is the TPM's vendor-rooted identity key, used to
certify that the AK lives in a genuine TPM (the basis for instance
identity). The \textbf{qualifying data / \texttt{ExtraData}} are
caller-supplied bytes folded into \texttt{TPMS\_ATTEST} and covered by
the quote's signature; this is exactly where a challenge nonce (and the
AEP's output-binding) is bound. The significance of \texttt{ExtraData}
for freshness is the crux of §6.

\hypertarget{eat-ar4si-ear}{%
\subsubsection{2.3 EAT, AR4SI, EAR}\label{eat-ar4si-ear}}

An \textbf{EAT (Entity Attestation Token)}, RFC 9711, is a standard
CWT/JWT-based token format for conveying Evidence and Attestation
Results as claims, with nesting via submodules. \textbf{AR4SI
(Attestation Results for Secure Interactions)} is an IETF Internet-Draft
(\texttt{draft-ietf-rats-ar4si}) {[}3{]} defining a trustworthiness
vocabulary: four tiers (\textbf{none / affirming / warning /
contraindicated}) across a vector of trust claims (e.g.,
\texttt{executables}, \texttt{configuration}). \textbf{EAR (EAT
Attestation Results)} is an IETF Internet-Draft
(\texttt{draft-ietf-rats-ear}) {[}4{]} serialising AR4SI results as a
signed EAT, carrying a per-submodule \texttt{ear.status} plus an
\texttt{ear.trustworthiness-vector}. In our runs the EAR profile is
\texttt{tag:github.com,2023:veraison/ear}, signed with ES256.

\hypertarget{corim-comid-cmw}{%
\subsubsection{2.4 CoRIM, CoMID, CMW}\label{corim-comid-cmw}}

A \textbf{CoRIM (Concise Reference Integrity Manifest)} is an IETF
Internet-Draft (\texttt{draft-ietf-rats-corim}) {[}5{]}: a CBOR manifest
that conveys Reference Values and trust anchors to a Verifier. A
\textbf{CoMID (Concise Module Identifier)} is the tag type inside a
CoRIM that actually carries the triples (e.g.,
\texttt{attester-verification-keys} for a trust anchor;
\texttt{reference-values} for golden measurements). A \textbf{CMW (RATS
Conceptual Message Wrapper)} is defined in RFC 9999 {[}6{]}: a wrapper
that groups multiple conceptual messages (e.g., platform Evidence plus
the AEP) into one message.

\hypertarget{aep-and-output-binding}{%
\subsubsection{2.5 AEP and
output-binding}\label{aep-and-output-binding}}

An \textbf{AEP (Action Evidence Package)} is a signed, append-only
record of what an agent did, who authorised it, the inputs and tool
calls, and the outcome, chained so that tampering shows {[}9{]}. The lab
preprint that seeded this family describes the same construct and is
candid about its limits: AEP-style packages can make selected agent
evidence inspectable and portable when producer and verifier share one
explicit package profile, but they do not, by themselves, prove truth,
runtime honesty, legal compliance, qualified trust-service status, or
policy wisdom; without a real runtime attester, an AEP does not prove
runtime honesty {[}12{]}. (The earlier lab material named the same
construct ``Agent Evidence Package,'' which we normalise to ``Action
Evidence Package'' in this article's prose.)

The bridge from ``signed record'' to ``attested record'' is
\textbf{output-binding}: in this implementation we measure
\texttt{H(AEP\ outcome)} into a PCR (PCR 4, §4.3), so the governance
outcome --- not a model identity --- is covered by the quote's signature
through the signed PCR composite, while the quote's qualifying data
carries the freshness nonce; tampering with the outcome changes the PCR
composite, and the quote no longer matches the golden reference (Figure
1). (Folding \texttt{H(outcome)\ ‖\ nonce} directly into the qualifying
data is the draft's alternative conveyance {[}8{]}, not exercised here.)
Finally, the \textbf{two-axis verdict} is the auditor-legible result: an
\textbf{authorisation axis} from the AEP (\textbf{Authorised /
Unauthorised / Indeterminate}) crossed with a \textbf{platform axis}
from the RATS appraisal (\textbf{Attested / Contested / Expired}). The
platform axis is what a conformant Verifier produces, and mapping
AR4SI/EAR onto it is the subject of §5.

\begin{center}\rule{0.5\linewidth}{0.5pt}\end{center}

\begin{figure}
\centering
\includegraphics[width=0.92\textwidth,height=\textheight]{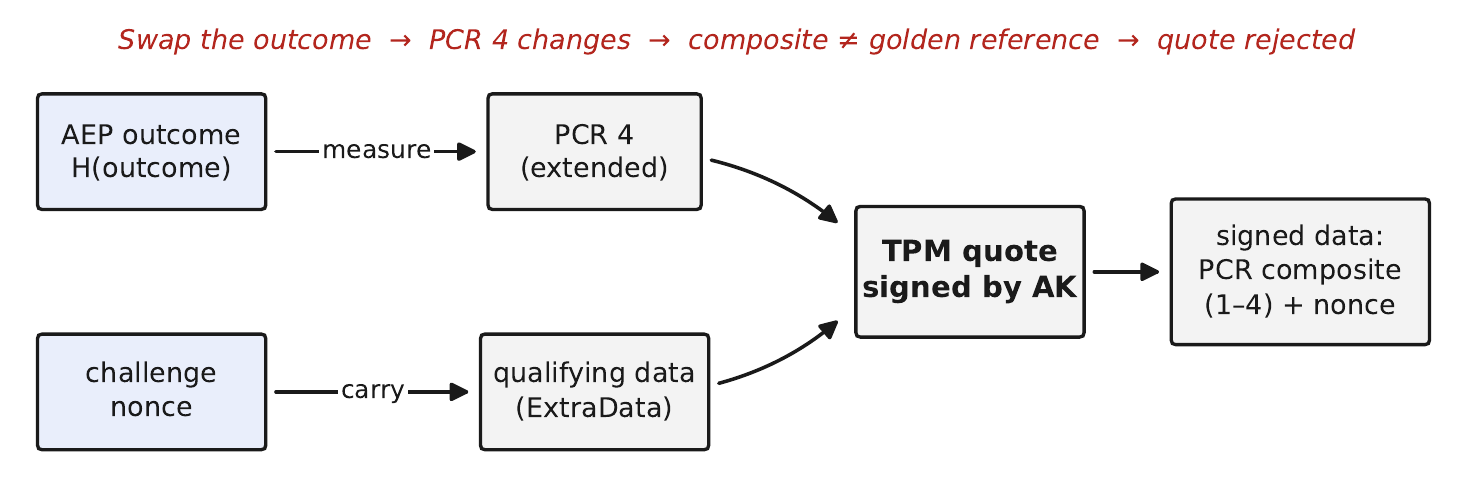}
\caption{Output-binding: the AEP outcome digest is measured into a PCR
that the quote signs over (the qualifying data carries the freshness
nonce), so swapping the outcome changes the signed PCR composite and the
quote no longer matches the golden reference.}
\end{figure}

\hypertarget{threat-model-and-scope}{%
\subsubsection{2.6 Threat model and
scope}\label{threat-model-and-scope}}

We state the model once, consolidated, and exercise its cases in
Sections 5 and 6.

\textbf{Assets.} The integrity of an agent's \emph{action evidence} (the
record of what an automated or automated agent did, who authorised it,
and the outcome) and the binding of that record to the platform state
that produced it.

\textbf{Adversary.} A party that controls the agent runtime and seeks to
evade accountability by crafting, altering, withholding, or replaying
evidence: for example, to make an unauthorised or harmful action appear
authorised and faithfully recorded, or to present a stale ``good'' state
as current. The adversary can author arbitrary AEPs, request quotes, and
capture and replay prior quote/evidence exchanges for an instance. The
adversary does \textbf{not} hold the platform Attestation Key's private
material and cannot extract it from a genuine root of trust.

\textbf{Trust boundaries.} The TPM and its Attestation Key (AK) are the
root: a quote's signature is assumed unforgeable without the AK private
key. The Verifier is trusted to appraise correctly against the
provisioned Reference Values and trust anchor; the
Reference-Value/Endorsement provider (the CoRIM author) is trusted for
the golden values; the Relying Party trusts the Verifier's signed EAR.

\textbf{In scope} (and exercised here): (i) \textbf{outcome
substitution}, swapping the AEP outcome under an otherwise-valid quote,
defeated by output-binding (Section 5, case B); (ii) \textbf{evidence
forgery}, tampering the signed quote structure, defeated by signature
verification (Section 5, case D); (iii) \textbf{replay/staleness},
re-presenting a previously valid quote as current (the freshness axis,
Section 6).

\textbf{Out of scope.} A compromised or counterfeit hardware root (AK
extraction, a subverted TPM); a compromised or colluding Verifier;
physical and side-channel attacks; the supply chain of the model or
agent code before its outcome is measured; and confidentiality of the
evidence. This work concerns integrity and freshness, not secrecy. The
demonstrated root is, moreover, an \textbf{emulated software TPM
(\texttt{swtpm}), not a hardware guarantee}, so the platform-root
assumption is exercised in emulation only (Sections 1.4 and 9.1).

\hypertarget{the-aep-on-rats-composition-recap}{%
\subsection{3. The AEP-on-RATS composition
(recap)}\label{the-aep-on-rats-composition-recap}}

The composition pattern this article instantiates is set out, as a
sketch for discussion, in the IETF Internet-Draft
\texttt{draft-sokolov-rats-aep-composition} (latest revision
\textbf{-02}; an \textbf{individual submission, NOT WG-adopted};
Informational; published on the datatracker 28 June 2026; expires 30
December 2026). We recap it here only insofar as the implementation
depends on it; the draft itself {[}8{]} is the authoritative statement,
which we summarise rather than reproduce.

The thesis is that an application-layer action record, the AEP, can be
treated as Evidence in the sense of the RATS Architecture (RFC 9334) and
bound to platform Evidence produced by a hardware root of trust, so that
a single Verifier, or a composition of Verifiers, can appraise both the
platform state and the application-layer action together and emit an
Attestation Result {[}8{]}. The draft is explicit that it is an
individual sketch, intended to ask the working group whether the pattern
is already covered by existing mechanisms or warrants a short document,
and we preserve that marker.

For conveyance, the draft names two candidates. The first is an EAT
{[}2{]} carrying the AEP (or a digest of it) as a claim or submodule,
using the EAT submodule / Detached-Submodule-Digest mechanism that is
the standard nesting facility here. The second is a CMW collection (the
RATS Conceptual Messages Wrapper {[}6{]}) that groups the platform
Evidence and the AEP into one message.

The security property the implementation must preserve is
output-binding. A forged AEP outcome presented under an otherwise-valid
platform quote MUST be detectable through it: the outcome digest is
covered by the quote's signed data, so an implementation that binds the
AEP reference \emph{outside} the signed data does not achieve the
property. The draft flags its own verdict mapping as provisional.
Expired is deliberately \emph{not} an AR4SI trustworthiness tier, and
the correspondence is provisional and SHOULD be validated against a
Verifier's actual EAR output, the validation we now perform. The draft
is equally clear that composition does not dissolve trust assumptions
but relocates them: the platform axis depends on the hardware vendor's
Endorsements and the Verifier's independence {[}8{]}.

In what follows we instantiate the platform axis concretely. We do
\emph{not} re-implement the full EAT-submodule/CMW conveyance for the
application-layer AEP in this article; instead we bind the AEP outcome
into the platform quote by measuring it into a PCR the quote signs over
(output-binding, §4.3) and drive that platform Evidence through
Veraison, which is sufficient to validate the platform axis of the
two-axis verdict and to surface the freshness finding. Full
multi-Verifier EAT-submodule composition is future work (§9).

\begin{center}\rule{0.5\linewidth}{0.5pt}\end{center}

\hypertarget{implementation}{%
\subsection{4. Implementation}\label{implementation}}

\begin{figure}
\centering
\includegraphics[width=0.98\textwidth,height=\textheight]{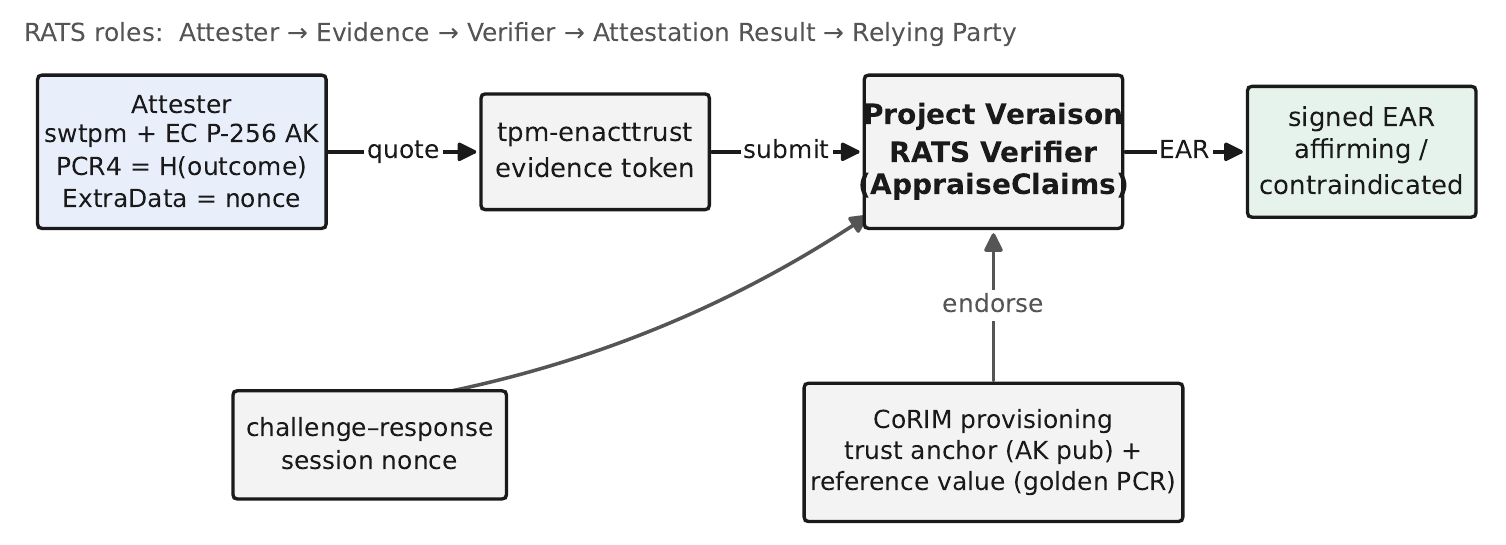}
\caption{The appraisal pipeline: an Attester (here \texttt{swtpm})
produces an output-bound quote that a CoRIM-provisioned Project Veraison
verifier appraises to a signed EAR; the node-id binds the trust anchor,
the reference value, and the submitted evidence.}
\end{figure}

All work in this section was executed on a single host (an Intel
i9-12900K workstation) on 27--28 June 2026. The pipeline, shown end to
end in Figure 2, is driven by one script, \texttt{aep-real-e2e.sh}:
\emph{swtpm EC AK → measure AEP outcome → provision CoRIM → quote →
submit → EAR}.

\hypertarget{the-attester-swtpm-with-an-ec-p-256-ak}{%
\subsubsection{4.1 The Attester: swtpm with an EC P-256
AK}\label{the-attester-swtpm-with-an-ec-p-256-ak}}

We run an emulated software TPM (\texttt{swtpm}) as the Attester. Inside
it we generate a fresh \textbf{EC P-256, ECDSA, SHA-256} attestation
key. This choice is not incidental: the \texttt{tpm-enacttrust} scheme
expects an ECDSA P-256 key, and an earlier RSA AK from the prior
\texttt{swtpm}-only demonstration had to be regenerated as EC for the
conformant path to accept it. We state plainly, and repeat, that this
Attester is an \textbf{emulated software TPM (\texttt{swtpm}), not a
hardware guarantee}.

\hypertarget{measuring-the-aep-outcome-into-pcr-4}{%
\subsubsection{4.2 Measuring the AEP outcome into PCR
4}\label{measuring-the-aep-outcome-into-pcr-4}}

The governance artifact is a demo AEP for a single governed tool call.
Its outcome digest (\texttt{outcome\_sha256}),

\begin{verbatim}
15dc47e8434e1b169fa0499a3b862db9f6238a1eacd61aee83c155a9e51c5743
\end{verbatim}

is extended into \textbf{PCR 4}. Because a PCR can only be extended
(\texttt{PCR\_new\ =\ H(PCR\_old\ ‖\ measurement)}), the value of PCR 4
now reflects the exact outcome that the agent recorded; any change to
the outcome changes PCR 4, and therefore the quote. This is the concrete
mechanism behind output-binding for the platform axis.

\hypertarget{producing-a-genuine-quote}{%
\subsubsection{4.3 Producing a genuine
quote}\label{producing-a-genuine-quote}}

We produce a genuine \texttt{swtpm} TPM quote over PCRs \textbf{1, 2, 3,
4} with the EC AK:

\begin{verbatim}
tpm2_quote -c $AKH -l sha256:1,2,3,4 -q $NONCE_HEX -m q.msg -s q.sig -g sha256
\end{verbatim}

PCR 4 carries \texttt{H(AEP\ outcome)}. The \textbf{golden PCR composite
digest} is \texttt{sha256(} concatenation of the four 32-byte PCR values
\texttt{)}, base64-encoded; this is the value provisioned as the
Verifier's reference. The session nonce is bound into the quote's
qualifying data via \texttt{-q\ \$NONCE\_HEX}; this fact is relevant
later because the \emph{scheme} (as opposed to our own appraiser) does
not look at it (§6).

For independent recomputation, the four PCRs are extended from the
\texttt{swtpm} reset state (all-zero) with one measurement each: PCR 1
with \texttt{sha256("firmware-v1")}, PCR 2 with
\texttt{sha256("bootloader-v1")}, PCR 3 with
\texttt{sha256("kernel-v1")}, and PCR 4 with the AEP outcome digest
\texttt{H(outcome)\ =\ 15dc47e8…51c5743}. Each register is therefore
\texttt{sha256(0}\^{}\texttt{32\ ‖\ measurement)}, and the golden
composite is \texttt{sha256(PCR1\ ‖\ PCR2\ ‖\ PCR3\ ‖\ PCR4)} =
\texttt{er5PQ51V3LpX2I6D3li4pZqG1CiGdQqAUJpeqzMWIiE=}. The
\texttt{B\_outcome\_swapped} case re-extends PCR 4 with
\texttt{sha256("a\ DIFFERENT\ fabricated\ summary")}, giving the
divergent composite
\texttt{cQhf0UzaSuKcvrRBmQVMqs0O/WLQAvvIC2+FXupbM10=}. Both digests are
reproducible from these inputs alone.

\hypertarget{packing-the-tpm-enacttrust-wire-format}{%
\subsubsection{4.4 Packing the tpm-enacttrust wire
format}\label{packing-the-tpm-enacttrust-wire-format}}

Veraison's \texttt{tpm-enacttrust} scheme {[}14{]} expects an evidence
token in a specific concatenated wire format:

\begin{verbatim}
NODE_ID (16 bytes, UUID) ‖ SIZE (uint16, big-endian, len of TPMS_ATTEST) ‖ TPMS_ATTEST ‖ TPMT_SIGNATURE
\end{verbatim}

The exact packer line (verbatim from \texttt{aep-real-e2e.sh}) is:

\begin{Shaded}
\begin{Highlighting}[]
\BuiltInTok{open}\NormalTok{(}\SpecialStringTok{f"}\SpecialCharTok{\{}\NormalTok{W}\SpecialCharTok{\}}\SpecialStringTok{/token.bin"}\NormalTok{,}\StringTok{"wb"}\NormalTok{).write(uuid.UUID(node).}\BuiltInTok{bytes} \OperatorTok{+}\NormalTok{ struct.pack(}\StringTok{\textquotesingle{}\textgreater{}H\textquotesingle{}}\NormalTok{,}\BuiltInTok{len}\NormalTok{(msg)) }\OperatorTok{+} \BuiltInTok{bytes}\NormalTok{(msg) }\OperatorTok{+}\NormalTok{ sig)}
\end{Highlighting}
\end{Shaded}

Here \texttt{NODE\_ID} is the instance UUID that keys the trust anchor
and reference value in the CoRIM; \texttt{TPMS\_ATTEST} is the
\texttt{tpm2\_quote} message (\texttt{q.msg}); and
\texttt{TPMT\_SIGNATURE} is the quote signature (\texttt{q.sig}). The
token is submitted with
\texttt{Content-Type:\ application/vnd.enacttrust.tpm-evidence}. The
instance UUID reported in the EAR for our run is
\texttt{345ccd98-a30b-4baf-9aa9-9b0861d2c042}.

\hypertarget{provisioning-a-corim-trust-anchor-reference-value}{%
\subsubsection{4.5 Provisioning a CoRIM (trust anchor + reference
value)}\label{provisioning-a-corim-trust-anchor-reference-value}}

We provision Veraison with a CoRIM carrying two CoMIDs:

\begin{enumerate}
\def\labelenumi{\arabic{enumi}.}
\tightlist
\item
  a \textbf{trust-anchor} CoMID: \texttt{attester-verification-keys},
  the AK public key as a \texttt{pkix-base64-key}, keyed by the instance
  UUID; and
\item
  a \textbf{reference-value} CoMID:
  \texttt{reference-values\ →\ measurements\ →\ digests:\ {[}"sha-256;\textless{}golden-b64\textgreater{}"{]}},
  keyed by the same instance UUID.
\end{enumerate}

Both are wrapped into a CoRIM with
\texttt{profile:\ https://enacttrust.com/veraison/1.0.0}, built with
\texttt{cocli\ comid\ create} then \texttt{cocli\ corim\ create}, and
submitted with \texttt{cocli\ corim\ submit}. The golden PCR composite
digest provisioned for the good run (the \texttt{A\_good\_fresh}
reference) is, base64-encoded over PCR selection 1,2,3,4 (SHA-256):

\begin{verbatim}
er5PQ51V3LpX2I6D3li4pZqG1CiGdQqAUJpeqzMWIiE=
\end{verbatim}

\hypertarget{challenge-response-to-an-ear}{%
\subsubsection{4.6 Challenge-response to an
EAR}\label{challenge-response-to-an-ear}}

A verification session is one challenge-response per case:
\texttt{POST\ newSession?nonceSize=32}, read the \texttt{Location:}
header (it is \textbf{relative}: prepend the
\texttt{challenge-response/v1} base URL) and the server \texttt{nonce},
submit the token to that session URL, and read back
\texttt{status:complete} with \texttt{result:} set to the EAR. The EAR
is a JWT; the payload is the second dot-segment. The Verifier-id
\texttt{developer} in our EARs is \texttt{Veraison\ Project}.

\hypertarget{the-gotchas-called-out-because-they-cost-real-time}{%
\subsubsection{4.7 The gotchas (called out because they cost real
time)}\label{the-gotchas-called-out-because-they-cost-real-time}}

These provisioning details are exact and were failure points; we surface
them so the artifact is reproducible without re-incurring the cost:

\begin{enumerate}
\def\labelenumi{\arabic{enumi}.}
\tightlist
\item
  \textbf{Media-type/profile must be exact.} Provisioning media-type
  must be
  \texttt{application/rim+cbor;\ profile="https://enacttrust.com/veraison/1.0.0"}.
  Using \texttt{corim-unsigned+cbor}, or the non-TLS \texttt{http://}
  profile string, returns \textbf{HTTP 415}.
\item
  \textbf{Instance binding must agree in three places.} The same
  \texttt{node-id} UUID must appear in (a) the trust-anchor CoMID, (b)
  the reference-value CoMID, and (c) the evidence token's leading 16
  bytes; otherwise the Verifier finds no anchor or reference for the
  instance.
\item
  \textbf{Relative session \texttt{Location}.} The session URL returned
  by \texttt{newSession} is relative and must be resolved against the
  challenge-response base.
\item
  \textbf{Schemes are go-plugins baked into the images.} Changing the
  scheme (e.g., the freshness patch of §6) requires
  \texttt{make\ docker-deploy} to rebuild the
  \texttt{vts}/\texttt{verification} images, not merely a container
  restart.
\item
  \textbf{\texttt{eat\_nonce} is not freshness enforcement.} The EAR
  echoes the session nonce as \texttt{eat\_nonce}, but the
  \texttt{tpm-enacttrust} scheme never compares it to the quote's
  \texttt{ExtraData}, the subject of §6.
\end{enumerate}

\hypertarget{buildrun-essentials}{%
\subsubsection{4.8 Build/run essentials}\label{buildrun-essentials}}

The Verifier is the upstream Project Veraison \texttt{services}
repository, built and run in Docker on the same host:
\texttt{git\ clone\ https://github.com/veraison/services.git};
\texttt{make\ docker-deploy} builds the \texttt{vts}, provisioning,
verification, management, keycloak and coserv images; and
\texttt{deployments/docker/veraison\ start} brings all six up. The
verification API listens on
\texttt{https://localhost:8443/challenge-response/v1} (self-signed TLS
in the dev deploy; \texttt{curl\ -k}). The complete driver is
\texttt{aep-real-e2e.sh}, and an independent re-verifier,
\texttt{diag-verify.go}, recomputes PCRDigest-versus-golden and checks
the signature.

\begin{center}\rule{0.5\linewidth}{0.5pt}\end{center}

\hypertarget{evaluation}{%
\subsection{5. Evaluation}\label{evaluation}}

We evaluate three cases that exercise the platform axis of the two-axis
verdict end-to-end: a good run, an outcome swap, and a signature tamper.
Each is a real \texttt{swtpm} quote appraised by the real Veraison
Verifier, and each result is read from a decoded EAR.

\hypertarget{the-verdict-mapping-table}{%
\subsubsection{5.1 The verdict-mapping
table}\label{the-verdict-mapping-table}}

Table 1 reports each case: the manipulation, the Veraison
\texttt{ear.status}, the \texttt{executables} trustworthiness tier from
the EAR \texttt{trustworthiness-vector}, and the released EAR file; §5.5
maps these onto the earlier \texttt{swtpm}-only verdicts. The case
labels A, B, and D are inherited from that earlier four-verdict
demonstration (A = Attested, B = Contested, C = Expired, D =
forged-rejected); case C (Expired/replay) is intentionally absent from
Table 1 because the reference scheme does not produce it, which is
precisely the freshness finding of Section 6.

\textbf{Table 1.} Cases A (\texttt{A\_good\_fresh}), B
(\texttt{B\_outcome\_swapped}), D (\texttt{D\_sig\_tampered}): real
\texttt{swtpm} AEP evidence appraised by real Veraison.

\begin{longtable}[]{@{}
  >{\raggedright\arraybackslash}p{(\columnwidth - 8\tabcolsep) * \real{0.0660}}
  >{\raggedright\arraybackslash}p{(\columnwidth - 8\tabcolsep) * \real{0.4623}}
  >{\raggedright\arraybackslash}p{(\columnwidth - 8\tabcolsep) * \real{0.1509}}
  >{\raggedright\arraybackslash}p{(\columnwidth - 8\tabcolsep) * \real{0.0660}}
  >{\raggedright\arraybackslash}p{(\columnwidth - 8\tabcolsep) * \real{0.2547}}@{}}
\toprule\noalign{}
\begin{minipage}[b]{\linewidth}\raggedright
Case
\end{minipage} & \begin{minipage}[b]{\linewidth}\raggedright
Manipulation
\end{minipage} & \begin{minipage}[b]{\linewidth}\raggedright
\texttt{ear.status}
\end{minipage} & \begin{minipage}[b]{\linewidth}\raggedright
Tier
\end{minipage} & \begin{minipage}[b]{\linewidth}\raggedright
EAR file
\end{minipage} \\
\midrule\noalign{}
\endhead
\bottomrule\noalign{}
\endlastfoot
A & good state; outcome measured into PCR 4 & \textbf{affirming} &
\texttt{2} & \texttt{ear-A\_good\_fresh.json} \\
B & PCR 4 re-measured with a different outcome, so the composite
diverges from golden & \textbf{contraindicated} & \texttt{33} &
\texttt{ear-B\_outcome\_swapped.json} \\
D & one byte flipped inside \texttt{TPMS\_ATTEST}, so the signature
fails & \textbf{contraindicated} & \texttt{99} &
\texttt{ear-D\_sig\_tampered.json} \\
\end{longtable}

The per-case EAR trustworthiness values, read directly from
\texttt{ear.trustworthiness-vector}, are \texttt{2} (affirming) for the
good run, \texttt{33} for the outcome-swapped run, and \texttt{99} for
the signature-tamper run, while each failing case carries submodule
\texttt{ear.status\ =\ contraindicated}. One nuance: under AR4SI's tier
numbering, \texttt{33} falls in the \emph{warning} band (32-95) rather
than the \emph{contraindicated} band (96-127). Veraison thus reports the
per-claim \texttt{executables} tier at \texttt{33} for the
reference-value mismatch yet sets the submodule status to
\texttt{contraindicated}; the \texttt{99} (whole-evidence failure) case
contraindicates every claim. We surface this as an observation to
confirm with the Veraison maintainers rather than asserting a clean tier
match; our own freshness policy (Section 6) uses the contraindicated
tier \texttt{96} explicitly.

The prior provisional mapping (\emph{Affirming → Attested; Warning or
Contraindicated → Contested}) is now confirmed against a real EAR for
the affirming and the reference-mismatch cases, and the forgery case
lands cleanly on \texttt{contraindicated} via signature failure.

\hypertarget{the-load-bearing-case}{%
\subsubsection{5.2 The load-bearing case}\label{the-load-bearing-case}}

The \texttt{B\_outcome\_swapped} case is the load-bearing one for the
AEP construct, just as it was in the prior feature's design {[}9{]}: a
perturbed model measurement surfaces as \textbf{Contested} --- the
\emph{(Authorised, Contested)} case the composite exists to expose ---
and the binding shows that an attacker cannot keep a valid quote while
swapping the outcome. What §5.1 adds is that this is no longer
demonstrated against a stand-in. Re-measuring PCR 4 with a different
outcome makes the quoted PCR composite diverge from the provisioned
golden reference; the conformant Verifier returns
\texttt{contraindicated} with \texttt{executables\ =\ 33}. The divergent
\texttt{pcr-digest} reported in the \texttt{B\_outcome\_swapped} EAR is:

\begin{verbatim}
cQhf0UzaSuKcvrRBmQVMqs0O/WLQAvvIC2+FXupbM10=
\end{verbatim}

distinct from the golden
\texttt{er5PQ51V3LpX2I6D3li4pZqG1CiGdQqAUJpeqzMWIiE=}. The
output-binding property holds: an attacker cannot keep a valid,
affirming quote while swapping the outcome, because the outcome is in
the signed PCR composite.

\hypertarget{the-verifiers-own-diagnosis-of-the-tamper}{%
\subsubsection{5.3 The Verifier's own diagnosis of the
tamper}\label{the-verifiers-own-diagnosis-of-the-tamper}}

For \texttt{D\_sig\_tampered} we flip one byte inside
\texttt{TPMS\_ATTEST} (\texttt{msg{[}-1{]}\ \^{}=\ 0x01}), which changes
\texttt{sha256(Raw)} so ECDSA verification fails. The Verifier does not
merely return \texttt{contraindicated}; it carries its own diagnosis in
the EAR \texttt{ear.veraison.policy-claims.problem} (verbatim):

\begin{quote}
``integrity validation failed: bad evidence: could not verify evidence
signature: failed to verify signature''
(\texttt{ear-D\_sig\_tampered.json})
\end{quote}

This is a meaningful difference from the stand-in: the conformant
Verifier produces structured, machine-readable provenance for the
failure, suitable for a Relying Party's audit trail.

\hypertarget{independent-re-verification}{%
\subsubsection{5.4 Independent
re-verification}\label{independent-re-verification}}

To guard against trusting the Verifier blindly, we independently
re-verify the \texttt{A\_good\_fresh} case with \texttt{diag-verify.go}.
It decodes the submitted token and checks two things directly: that the
quote's \texttt{PCRDigest} \textbf{equals} the provisioned golden, and
that the signature \textbf{verifies} against our AK public key, which is
exactly what Veraison checks. Both hold. The independent re-verification
gives us confidence that the affirming verdict reflects the evidence,
not a misconfiguration of the Verifier.

\hypertarget{relation-to-the-four-earlier-verdicts}{%
\subsubsection{5.5 Relation to the four earlier
verdicts}\label{relation-to-the-four-earlier-verdicts}}

The earlier \texttt{swtpm}-only demonstration produced \textbf{four}
verdicts (Attested, Contested, Expired, and forged-rejected) for AEP
digest
\texttt{b80c48d9115da3ed8d388f670af5d713cfd990d6743c77094495fcabe3e59d89},
one sub-second run per case. Against the conformant Veraison
\texttt{tpm-enacttrust} scheme, \textbf{three of those four} map onto
its affirming/contraindicated outcomes: Attested to affirming, Contested
to contraindicated (reference-value divergence), and forged-rejected to
contraindicated (signature failure). The \textbf{Expired/freshness}
verdict, however, is \emph{not} produced by the \emph{unpatched}
Veraison scheme. That absence is not a curiosity; it is a security
finding, and §6 closes it with a validated fix.

\begin{center}\rule{0.5\linewidth}{0.5pt}\end{center}

\hypertarget{a-freshness-gap-in-a-deployed-verifier}{%
\subsection{6. A freshness gap in a deployed
verifier}\label{a-freshness-gap-in-a-deployed-verifier}}

\hypertarget{the-finding}{%
\subsubsection{6.1 The finding}\label{the-finding}}

While mapping the earlier verdicts onto the Veraison appraisal we found
that the \textbf{Expired/replay} verdict had no counterpart in the
conformant path. Investigating, we confirmed a security-relevant gap in
the reference scheme. Stated precisely: Veraison's reference
\texttt{tpm-enacttrust} scheme appraises the \textbf{signature} and the
\textbf{PCR digest versus the reference value only}; it does
\textbf{not} enforce challenge-nonce freshness. The scheme never
compares the quote's qualifying data (\texttt{TPMS\_ATTEST.ExtraData},
where the attester binds the challenge nonce) against the session's
expected nonce, so a replayed or stale quote still returns
\texttt{affirming}. A source check confirms this: a \texttt{grep} for
\texttt{nonce} or \texttt{extradata} across
\texttt{scheme/tpm-enacttrust/*.go} returns nothing, and the EAR's
\texttt{eat\_nonce} is merely the framework echoing the session nonce.

In words: the verification framework issues a per-session nonce and
echoes it back in the EAR as \texttt{eat\_nonce}, which \emph{looks}
like a freshness check. But the \texttt{tpm-enacttrust} scheme (the
component that actually appraises the evidence) never reads the quote's
\texttt{ExtraData}, where the Attester binds that nonce into the signed
data. So nothing ties the appraised evidence to the live session. A
captured quote, replayed later (or an old quote re-submitted), still
satisfies the scheme's two checks (the signature is still valid, the PCR
composite still matches the golden reference) and the Verifier returns
\texttt{affirming}. The \texttt{eat\_nonce} echo gives a false sense of
freshness: it reflects the session the framework created, not a property
the scheme verified against the evidence.

The practical consequence is that, in the \emph{unpatched} scheme, the
earlier demonstration's Expired/replay verdict cannot be produced by
Veraison; that check lived only in our own minimal appraiser, which does
compute it
(\texttt{tpm2\_checkquote\ -q\ \textless{}nonce\textgreater{}} rejects a
quote whose qualifying data does not match the expected nonce), until we
add it to the scheme itself and validate the flip in §6.3. Nonce
freshness is not a novel requirement. It is one of Coker et al.'s five
principles of remote attestation {[}17{]}, and formal analyses have
repeatedly surfaced replay or diversion gaps in attestation protocols,
including RA-TLS {[}18{]} and EPID-based SGX attestation {[}19{]}, and
self-measurement schemes for unattended devices add explicit
temporal-consistency {[}29{]} and time-of-check-to-time-of-use {[}30{]}
defences against replay-evasion. The pattern here is freshness
\emph{delegated to the wrong layer}: RFC 9334's Section 10 names nonces,
synchronised-clock timestamps, and epoch IDs as the freshness mechanisms
{[}1{]}, and Veraison's challenge-response transport does supply a
per-session nonce, but enforcement is left to the appraisal scheme,
where the two TPM schemes silently drop it while the platform schemes
keep it. Our contribution is therefore not to rediscover that freshness
matters, but to show that a \emph{deployed, conformant reference scheme}
silently omits the check while echoing a nonce that masks its absence
--- and to give the exact two-part fix.

\textbf{The omission is not unique to \texttt{tpm-enacttrust}.} A source
audit of the appraisal path
(\texttt{AppraiseClaims}/\texttt{ExtractClaims} in
\texttt{scheme/\textless{}name\textgreater{}/scheme.go}) of the schemes
shipped in the current Veraison tree shows the gap is a scheme-family
pattern. Both of Veraison's TPM-based schemes skip the check; every
platform-attestation scheme audited enforces it in-scheme, each with an
explicit, labelled \emph{freshness} comparison that rejects a mismatched
nonce.

\begin{longtable}[]{@{}
  >{\raggedright\arraybackslash}p{(\columnwidth - 4\tabcolsep) * \real{0.1758}}
  >{\raggedright\arraybackslash}p{(\columnwidth - 4\tabcolsep) * \real{0.1099}}
  >{\raggedright\arraybackslash}p{(\columnwidth - 4\tabcolsep) * \real{0.7143}}@{}}
\toprule\noalign{}
\begin{minipage}[b]{\linewidth}\raggedright
Scheme
\end{minipage} & \begin{minipage}[b]{\linewidth}\raggedright
Type
\end{minipage} & \begin{minipage}[b]{\linewidth}\raggedright
Enforces challenge-nonce freshness?
\end{minipage} \\
\midrule\noalign{}
\endhead
\bottomrule\noalign{}
\endlastfoot
\texttt{psa-iot} & platform & Yes: compares the token \texttt{psa-nonce}
to the session nonce \\
\texttt{arm-cca} & platform & Yes: compares the realm challenge to the
session nonce \\
\texttt{sevsnp} & platform & Yes: \texttt{validateSessionNonce} compares
report data to the session nonce \\
\texttt{tpm-enacttrust} & TPM & \textbf{No}: appraises signature and PCR
digest only; the nonce is never compared \\
\texttt{parsec-tpm} & TPM & \textbf{No}: appraises signature and PCR
digest only; the nonce is never compared \\
\end{longtable}

The framework delegates freshness to the scheme; the two TPM schemes
forget it, while \texttt{psa-iot}, \texttt{arm-cca}, and \texttt{sevsnp}
supply the exact template the fix of §6.3 ports to
\texttt{tpm-enacttrust}. This generalises the finding from a single
scheme to a TPM-scheme-family pattern and shows the fix is idiomatic
rather than novel.

\hypertarget{threat-model-and-impact}{%
\subsubsection{6.2 Threat model and
impact}\label{threat-model-and-impact}}

The output-binding threat model (§3) requires that a forged outcome
under an otherwise-valid quote be detectable. The freshness gap is a
\emph{different} axis: it concerns whether the quote is \emph{current}.
An adversary who can capture a single affirming quote-and-token for an
instance (e.g., by observing one legitimate exchange) can replay it to
obtain \texttt{affirming} for that instance at any later time,
regardless of the instance's current state, as long as the provisioned
reference and key are unchanged. For an automated-agent accountability
use case this is consequential: a Relying Party that treats an
\texttt{affirming} EAR as evidence that the agent's runtime is
\emph{now} in a good state would be misled by a stale quote. Freshness
is not a nicety here; it is part of what makes attestation evidence of
present state rather than of a past moment.

\hypertarget{the-fix-exact-idiomatic-two-part}{%
\subsubsection{6.3 The fix (exact, idiomatic,
two-part)}\label{the-fix-exact-idiomatic-two-part}}

The root cause is structural, not a deep design flaw: the scheme decodes
the token (which \emph{does} contain \texttt{ExtraData}) in
\texttt{ExtractClaims}, and the OPA policy engine already receives
\texttt{input.session} (the verification context). The missing link is
that the scheme never puts the attester-bound nonce into the claims, so
any policy has nothing to compare against. The idiomatic Veraison fix is
therefore two small pieces.

\textbf{(1) Scheme patch: surface the attester nonce as a claim.} In
\texttt{scheme/tpm-enacttrust/scheme.go}, in \texttt{ExtractClaims},
immediately after the \texttt{pcr-digest} line (\textasciitilde line
72):

\begin{Shaded}
\begin{Highlighting}[]
\NormalTok{    claims["pcr{-}digest"] = []byte(decoded.AttestationData.AttestedQuoteInfo.PCRDigest)}
\VariableTok{+   // Surface the attester{-}bound qualifying data (TPMS\_ATTEST.ExtraData) so an}
\VariableTok{+   // appraisal policy can enforce challenge{-}nonce freshness against input.session.}
\VariableTok{+   claims["nonce"] = []byte(decoded.AttestationData.ExtraData)}
 
\NormalTok{    return claims, nil}
\end{Highlighting}
\end{Shaded}

Because schemes are go-plugins baked into the images, the plugin must be
rebuilt and redeployed with \texttt{make\ docker-deploy} (use a clean
rebuild, \texttt{make\ really-clean\ \&\&\ make\ docker-deploy}, since a
plain rebuild cache-hits the scheme compile).

\textbf{(2) Appraisal policy: enforce freshness (Rego).} With the nonce
now a claim, a small Rego policy compares the attester-bound nonce to
the session's expected nonce, normalising the two base64 encodings the
framework uses (the evidence nonce is standard base64, the session nonce
URL-safe), and overrides the submodule \textbf{status} to
\texttt{contraindicated} on mismatch, passing the scheme's own verdict
through on a match:

\begin{Shaded}
\begin{Highlighting}[]
\NormalTok{package policy}
\NormalTok{import future.keywords}
\NormalTok{tier := \{"none": 0, "affirming": 2, "warning": 32, "contraindicated": 96\}}
\NormalTok{\# input.evidence.nonce : attester{-}bound nonce (TPMS\_ATTEST.ExtraData), standard base64}
\NormalTok{\# input.session.nonce  : session\textquotesingle{}s expected nonce (EAT eat\_nonce), URL{-}safe base64}
\NormalTok{session\_std := replace(replace(input.session.nonce, "{-}", "+"), "\_", "/")}
\NormalTok{fresh := input.evidence.nonce == session\_std}
\NormalTok{status := tier[input.result["ear.status"]] \{ fresh \}   \# fresh: pass the scheme verdict through}
\NormalTok{status := 96 \{ not fresh \}                              \# replay/stale: contraindicated}
\end{Highlighting}
\end{Shaded}

The policy overrides only the submodule \texttt{ear.status} (the
released \texttt{ear-REPLAY.json} accordingly shows
\texttt{ear.status\ =\ contraindicated} with the per-claim
\texttt{executables} tier unchanged). It is created and activated with
\texttt{pocli}. The management API is HTTPS and policy creation requires
the manager role:
\texttt{pocli\ create\ TPM\_ENACTTRUST\ freshness-policy.rego\ -s\ -i\ -U\ veraison-manager}.

\textbf{Validation, measured.} We executed this fix end-to-end on a
patched build. With the scheme patch surfacing the attester-bound nonce
as a claim and the Rego policy above active, the \emph{same} valid,
correctly-signed \texttt{swtpm} AEP quote appraises differently
according to freshness (Figure 3): submitted to its own
challenge-response session it returns \texttt{affirming}; replayed to a
fresh session whose expected nonce differs, it returns
\texttt{contraindicated}. Both Attestation Results carry the same
\texttt{ear.appraisal-policy-id} and the same attester-bound
\texttt{nonce} claim (the replayed token is byte-identical) and differ
only in the session's expected nonce (\texttt{D/YLjUr5…} bound in the
quote, versus \texttt{VKvxfur0…} expected by the replay session). The
two EARs (\texttt{ear-FRESH.json}, \texttt{ear-REPLAY.json}) and the
driver that reproduces them (\texttt{flip-e2e.sh}) are released with the
artifact. The flip ran as a separate Verifier instance (node-id
\texttt{171a2b77-48c8-4ebd-b554-acaa599c7983}; the §5 verdict-mapping
run used \texttt{345ccd98-a30b-4baf-9aa9-9b0861d2c042}), each
provisioned with its own trust anchor and golden reference. This
measured \texttt{affirming\ →\ contraindicated} flip turns the finding
from \emph{reported gap plus proposed fix} into \emph{gap plus validated
fix}.

\begin{figure}
\centering
\includegraphics[width=0.92\textwidth,height=\textheight]{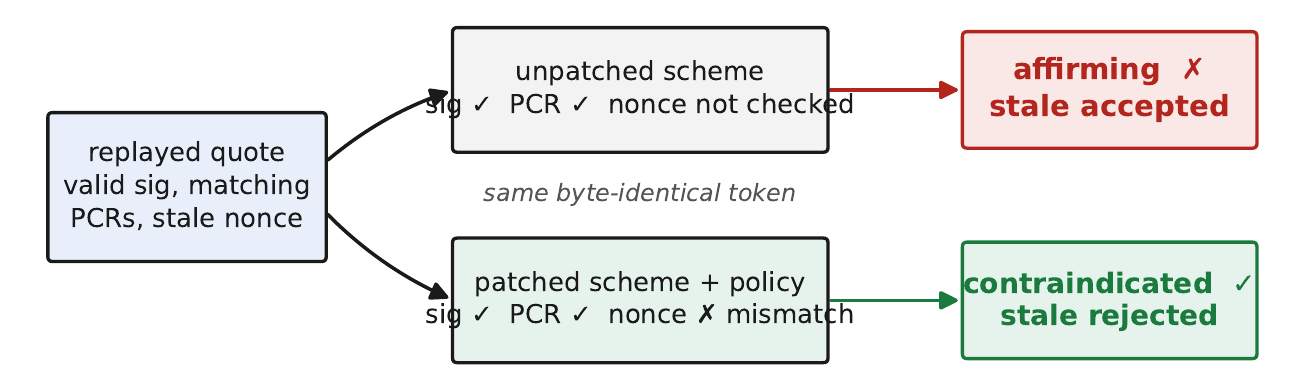}
\caption{The freshness flip. The same byte-identical replayed quote
appraises as \texttt{affirming} under the unpatched scheme (which checks
the signature and the PCR digest but never the nonce) and as
\texttt{contraindicated} once the scheme patch and freshness policy are
active, because the attester-bound nonce no longer matches the replay
session's expected nonce.}
\end{figure}

\hypertarget{status-of-the-fix-and-responsible-disclosure}{%
\subsubsection{6.4 Status of the fix, and responsible
disclosure}\label{status-of-the-fix-and-responsible-disclosure}}

On how finished the fix is: the one-line scheme patch and the Rego
policy are exact and validated end-to-end (§6.3), the policy normalises
the two base64 encodings the framework uses: the attester-bound nonce
arrives in standard base64 (the scheme surfaces a Go byte string), the
session's expected nonce in URL-safe base64 (the EAT
\texttt{eat\_nonce}), before the byte-for-byte compare, and passes the
scheme's own verdict through on a match. Settling the precise
policy-override semantics for upstreaming remains a conversation for the
Veraison maintainers (Fossati/Deshpande). We frame this finding in the
responsible-disclosure posture appropriate to a community-maintained
open-source project: the issue is a missing check in a \emph{reference}
scheme (not a deployed production trust service); the fix is small,
idiomatic, and upstreamable (``\texttt{tpm-enacttrust} should enforce
freshness; here's the one-line claim plus the policy'') and, consistent
with responsible disclosure, we reported it to the upstream Veraison
maintainers before any public deposit of this article or its artifacts
(\texttt{veraison/services} issue
\href{https://github.com/veraison/services/issues/427}{\#427}, filed 28
June 2026, ahead of the 28 June artifact deposit), as a missing check to
be improved rather than a latent zero-day. The disclosure has since been
acted on upstream: pull request
\href{https://github.com/veraison/services/pull/432}{\#432}, opened 10
July 2026 by a Veraison contributor and approved by a maintainer,
surfaces the attester-bound nonce as a claim and adds regression and
integration tests covering the replay case, including the base64 /
base64url encoding distinction noted below; the Rego policy comparison
of part 2 is not included there. We note explicitly that this
is a property of the \texttt{tpm-enacttrust} \emph{reference scheme} as
deployed; other Veraison schemes and policies may handle freshness
differently, and the framework's session machinery already carries the
nonce the fix needs.

\begin{center}\rule{0.5\linewidth}{0.5pt}\end{center}

\hypertarget{reproducibility-and-artifact}{%
\subsection{7. Reproducibility and
artifact}\label{reproducibility-and-artifact}}

The entire pipeline is reproducible from a single driver and a small set
of result files; we release them so an independent reader can re-run the
appraisal and re-derive every number in §5.

\textbf{Driver.} \texttt{aep-real-e2e.sh} runs the full pipeline:
\texttt{swtpm} EC AK generation → measure the AEP outcome into PCR 4 →
build and submit the CoRIM → quote → pack the \texttt{tpm-enacttrust}
token → submit to a fresh challenge-response session → decode the EAR.
It assumes a running Veraison deployment (§4.8).

\textbf{Result artifacts.} For each case we release the decoded EAR
(\texttt{ear-A\_good\_fresh.json},
\texttt{ear-B\_outcome\_swapped.json},
\texttt{ear-D\_sig\_tampered.json}) and the exact evidence token
submitted (\texttt{token-A\_good\_fresh.bin},
\texttt{token-B\_outcome\_swapped.bin},
\texttt{token-D\_sig\_tampered.bin}). The independent re-verifier
\texttt{diag-verify.go} recomputes PCRDigest-versus-golden and checks
the signature against the AK public key, so the affirming verdict can be
confirmed without trusting the Verifier. For the freshness fix (§6.3) we
additionally release the appraisal policy
(\texttt{freshness-policy.rego}), the driver \texttt{flip-e2e.sh}, and
the two flip EARs (\texttt{ear-FRESH.json}, \texttt{ear-REPLAY.json})
over the byte-identical replayed token, which together reproduce the
\texttt{affirming\ →\ contraindicated} flip on a patched build.

\textbf{Key identifiers} (so a reader can confirm they are reproducing
the same run): AEP outcome digest
\texttt{15dc47e8434e1b169fa0499a3b862db9f6238a1eacd61aee83c155a9e51c5743};
instance node-id \texttt{345ccd98-a30b-4baf-9aa9-9b0861d2c042}; golden
PCR digest \texttt{er5PQ51V3LpX2I6D3li4pZqG1CiGdQqAUJpeqzMWIiE=} (PCR
selection 1,2,3,4; SHA-256); EAR profile
\texttt{tag:github.com,2023:veraison/ear} (ES256). An earlier
reference-vector milestone (the first proof of the loop, before our own
quote was wired in) used node-id
\texttt{7df7714e-aa04-4638-bcbf-434b1dd720f1} and pcr-digest
\texttt{h0KPxSKAPTEGXnvOPPA/5HUJZjHl4Hu9eg/eYMTPJcc=}.

\textbf{Reproducibility lineage.} The conformance posture continues a
line from the AEP attestation lab, whose negative-control vectors
establish the producer/verifier contract this article's evidence tokens
implicitly satisfy: that lab reported a clean conformance result of 11
total vectors (4 expected-valid and 7 expected-rejected cases) with 11
contract-met cases and 0 contract mismatches {[}12{]}. Where the lab
established the package-level contract with positive and negative
vectors, this article carries a \emph{single, real} package through a
\emph{conformant Verifier}; the two are complementary layers of the same
reproducibility story.

\hypertarget{data-and-code-availability}{%
\subsubsection{7.1 Data and code
availability}\label{data-and-code-availability}}

The artifacts named above (the driver \texttt{aep-real-e2e.sh}, the
three decoded EARs, the submitted evidence tokens, the independent
re-verifier \texttt{diag-verify.go}, and the freshness-fix set:
\texttt{freshness-policy.rego}, \texttt{flip-e2e.sh}, the two flip EARs
\texttt{ear-FRESH.json}/\texttt{ear-REPLAY.json}, and the
scheme-freshness survey \texttt{scheme-freshness-survey.md}) are
released with this submission as a versioned, citable Zenodo deposit,
DOI 10.5281/zenodo.20998730 (CC-BY); the related preprint of the prior
IEEE Internet Computing work is at concept DOI 10.5281/zenodo.20818671.
The
Verifier is the open-source Project Veraison
(\texttt{github.com/veraison/services}), deployed through its Docker
flow (\texttt{make\ docker-deploy}). The reported run used
\texttt{swtpm} 0.7.3 (native; 0.7.1 in the container build),
\texttt{tpm2-tools} 5.6 (native; 5.4 in the container), and Docker
29.1.3 with Buildx 0.30.1, on a single Linux x86-64 host (Ubuntu 24.04).
The Attestation Key is EC P-256 (ECDSA / SHA-256) and the EAR is ES256.
No randomised split or seed applies: the pipeline is deterministic given
the measured PCR state and the per-session nonce.

\begin{center}\rule{0.5\linewidth}{0.5pt}\end{center}

\hypertarget{related-work}{%
\subsection{8. Related work}\label{related-work}}

\hypertarget{identity-versus-action-evidence}{%
\subsubsection{8.1 Identity versus
action-evidence}\label{identity-versus-action-evidence}}

A large body of attestation work concerns proving \emph{what a platform
is}: secure boot, measured boot, confidential computing, and the RATS
architecture itself (RFC 9334) with its token (EAT, RFC 9711) and result
(AR4SI, EAR) vocabularies. The contribution of the AEP line, and of this
article, is orthogonal to platform identity: it concerns \emph{what an
action was}, binding a governance outcome into the platform's signed
measurement so that the outcome inherits the platform's tamper-evidence.
The distinction matters because, for agent accountability, knowing that
a genuine TPM signed the record does not tell a reviewer whether the
recorded outcome is the one that actually occurred; output-binding is
what closes that. This builds on a long lineage of TPM-based remote
attestation, integrity measurement {[}15{]}, continuous TPM 2.0
attestation {[}16{]}, and the foundational principles of attestation
(including temporal freshness) {[}17{]}, surveyed more broadly in
{[}20{]}, {[}21{]}. It is also the hardware-rooted counterpart of recent
governance work on agent \emph{visibility} and \emph{infrastructure}
{[}22{]}, {[}23{]}, which calls for activity logging and behaviour
certification without grounding either in a hardware root.

\hypertarget{agent-action-and-execution-evidence-proposals}{%
\subsubsection{8.2 Agent-action and execution-evidence
proposals}\label{agent-action-and-execution-evidence-proposals}}

Several recent proposals carry assurances about agent actions or
executions. The lab preprint surveyed in this article's lineage situates
AEP relative to attestation vocabulary and is candid that AEP ``does not
replace'' RATS/EAT but reuses their roles (quoted in §2.5). More
broadly, the design space includes permit/receipt patterns that
authorise an effect \emph{before} it commits {[}24{]}: PermitReceipt
stops at the issuer's signature over an action digest and has no
hardware-state axis, so unlike AEP it never folds the action digest into
a quote's qualifying data nor proves what the runtime \emph{was} when it
acted. It includes proposals for verifiable agent conversations in the
RATS community {[}28{]}, which record interaction \emph{content} rather
than binding a recorded \emph{outcome} into platform Evidence; and
Sato's SOOS framing {[}26{]}, {[}27{]}, whose KIA/GAR sign
hardware-anchored audit records and emit governance labels (e.g.,
\texttt{ALLOW}/\texttt{DENY}/\texttt{ESCALATE}) but neither fold the
action record into a fresh quote's qualifying data nor produce a single
composed AR4SI/EAR verdict fusing a platform-trust axis with the
authority axis --- which is exactly AEP's contribution. (Sato's draft
uses the unrelated term \emph{Agent Execution Protocol}, not to be
conflated with our \emph{Action Evidence Package}.) Attested-TLS binds a
\emph{channel} to a platform attestation via exported keying material
{[}25{]}: where it binds a \emph{session} to a platform state,
output-binding binds a \emph{recorded outcome} to one, and the two
compose. This article does not claim to subsume these; it claims,
specifically, to be the first we are aware of to carry an
application-layer action-evidence construct through a \emph{conformant}
RATS Verifier to a signed EAR, and to surface a concrete freshness
finding in the reference scheme.

\hypertarget{the-authors-own-companion-surfaces-openly-disclosed}{%
\subsubsection{8.3 The authors' own companion surfaces (openly
disclosed)}\label{the-authors-own-companion-surfaces-openly-disclosed}}

In the interest of avoiding self-overlap, we disclose the author's own
related surfaces and state plainly how this article differs from each:

\begin{itemize}
\tightlist
\item
  \textbf{IEEE Internet Computing}, \emph{``When the Witness Is Also the
  Suspect: Hardware-Rooting the Evidence That AI Agents Leave Behind''}
  --- \textbf{under review} since 25 June 2026. This magazine feature
  motivates the problem and demonstrates output-binding against a
  \textbf{minimal Verifier stand-in} (\texttt{tpm2\_checkquote} plus a
  reference-PCR comparison). \emph{This article replaces that stand-in
  with a conformant Veraison Verifier}, precisely the gap the feature
  names as future work, and adds the freshness finding and fix.
\item
  \textbf{IEEE Transactions on Technology \& Society} {[}10{]},
  \emph{``Evidence Instrumentation for AI-Governance Review: Three
  Public-Source Practitioner Use Cases''} (MS 2026-06-0114-OTW-TTS) ---
  submitted 13 June 2026, \textbf{rejected upon initial review 18 July
  2026}. A companion, applied use-cases article; the term AEP is absent
  there. It does not overlap with the systems contribution of this
  article. Its underlying data remain publicly available (DOI
  10.5281/zenodo.20488643). A substantially reconstructed successor,
  \emph{``Claim-Evidence-Boundary Records for Public-Source AI-Governance
  Review''} (MS 2026-07-0152-RES-TTS), was submitted to the same journal
  on 23 July 2026; it likewise does not overlap with the systems
  contribution here.
\item
  \textbf{IETF Internet-Draft}
  \texttt{draft-sokolov-rats-aep-composition} (latest revision
  \textbf{-02}) --- an \textbf{individual submission, NOT WG-adopted},
  published on the datatracker 28 June 2026. It states the composition
  pattern and the provisional verdict mapping; this article performs the
  validation against a real EAR that the draft explicitly asks for.
\item
  \textbf{Zenodo preprint} {[}11{]}, DOI
  \textbf{10.5281/zenodo.20818671} (CC-BY, deposited 24 June 2026; current
  version 3, 31 July 2026) ---
  the public preprint of the IC manuscript.
\item
  \textbf{AEP attestation lab preprint v0.4}, \emph{``Policy-Bound Agent
  Evidence Packages''} (27 May 2026) --- reproducible
  conformance/negative-control vectors; uses the earlier ``Agent
  Evidence Package'' term for the same construct family.
\end{itemize}

This article quotes short, attributed excerpts from those surfaces
(above and throughout) rather than reusing their prose, and its distinct
contribution --- the conformant-verifier integration, the validated
verdict mapping, the freshness finding and its validated two-part fix,
and the reproducible artifact --- does not appear in any of them.

\begin{center}\rule{0.5\linewidth}{0.5pt}\end{center}

\hypertarget{limitations-and-future-work}{%
\subsection{9. Limitations and future
work}\label{limitations-and-future-work}}

\hypertarget{limitations}{%
\subsubsection{9.1 Limitations}\label{limitations}}

\textbf{Emulated, not hardware, root of trust.} The Attester is an
\textbf{emulated software TPM (\texttt{swtpm}), not a hardware
guarantee}. Nothing in this article should be read as a hardware-rooted
security claim; what we demonstrate is the \emph{appraisal path}
end-to-end, with the platform root deliberately emulated. A real
discrete STMicroelectronics dTPM run is staged on a laptop with a
hardware TPM 2.0 and will be carried out as a proof-of-concept at the
IETF 126 Hackathon (Vienna, 18--19 July 2026), alongside the Veraison
maintainers, to exercise the hardware-rooted guarantee that the emulated
root stands in for. The freshness finding and its fix (§6) are, however,
properties of the scheme's appraisal logic: whether the root is emulated
or hardware, \texttt{tpm-enacttrust} reads the same claims and skips the
same nonce check, so the security result itself does not depend on the
emulated root. The hardware run strengthens the output-binding trust
claim, not the finding.

\textbf{One scheme exercised end-to-end; the rest read.} We exercise
Veraison's reference \texttt{tpm-enacttrust} scheme end-to-end, and the
validated freshness flip (§6.3) is on that scheme. The wider result of
§6.1 is a \emph{source audit} of the schemes shipped in the current
Veraison tree, not an end-to-end run of each: the \texttt{parsec-tpm}
co-finding in particular is read from the code, not executed, and the
platform schemes' freshness checks are read from their
\texttt{AppraiseClaims}. Other schemes, policies, or future revisions
may behave differently.

\textbf{A single Verifier.} We use one Verifier instance (Project
Veraison). The composition pattern (§3) envisages a single Verifier
\emph{or a composition of Verifiers}; we validate only the
single-Verifier, platform-axis case, and we bind the AEP via
output-binding rather than implementing the full EAT-submodule/CMW
conveyance of the application-layer AEP.

\textbf{Platform axis only, in this article.} We validate the platform
axis (Attested/Contested/Expired) of the two-axis verdict against a real
EAR. The authorisation axis (Authorised/Unauthorised/Indeterminate)
comes from the AEP itself and is not re-derived through Veraison here;
and the Veraison scheme produces the Expired term once our freshness fix
is active (§6.3), which is exactly the gap that fix closes.

\hypertarget{future-work}{%
\subsubsection{9.2 Future work}\label{future-work}}

\textbf{A real discrete TPM.} Executing the dTPM run already set up
would replace the emulated root with a hardware one and let us state a
hardware-rooted claim, retiring the principal scope marker of this
article.

\textbf{Multi-Verifier, EAT-submodule composition.} Implementing the
application-layer AEP as an EAT submodule (or a CMW collection) and
appraising platform Evidence and the AEP across a composition of
Verifiers would validate the full composition the Internet-Draft
sketches, not just the platform axis.

\textbf{Upstreaming the freshness fix.} Our patched build already
produces the Expired term via the freshness policy (§6.3); finalising
the policy-override semantics with the Veraison maintainers and
contributing the two-part fix upstream would turn this validated fix
into a durable, native improvement to the reference scheme rather than a
local patch.

\begin{center}\rule{0.5\linewidth}{0.5pt}\end{center}

\hypertarget{conclusion}{%
\subsection{10. Conclusion}\label{conclusion}}

The accountability gap for automated-agent actions is not that agents
fail to keep records. It is that the records are kept by the very
runtime whose honesty is in question: the witness is also the suspect.
Remote attestation answers this by separating the party that produces
evidence from the party that judges it, and the Action Evidence Package
construct extends that separation to the application layer by binding a
governance outcome into a hardware-rooted quote. Prior work demonstrated
the binding but stopped at a minimal appraiser stand-in.

This article carries the construct the rest of the way. A real
\texttt{swtpm} AEP quote is provisioned and appraised end-to-end by a
conformant Project Veraison RATS Verifier, yielding a signed EAR:
affirming for good evidence, contraindicated for an outcome swap, and
contraindicated with a structured signature-failure diagnosis for a
tamper, independently re-verified. In doing so we validated the
previously-provisional verdict mapping against a real EAR. We also
found, characterised, and fixed a freshness gap in the deployed
\texttt{tpm-enacttrust} scheme (and, by a source audit, in
\texttt{parsec-tpm} too), giving an exact, idiomatic,
responsibly-disclosed two-part fix that we validate end-to-end: with it
active, a replayed quote flips from \texttt{affirming} to
\texttt{contraindicated}.

The Attester remains an emulated software TPM, not a hardware guarantee,
and the Internet-Draft underpinning the composition remains an
individual submission, not WG-adopted. We keep those markers explicit.
What changed is concrete and reproducible: the appraiser is now a
conformant Verifier, and the evidence an automated agent leaves behind
can be carried, intact and falsifiable, all the way to a
standards-conformant Attestation Result.

\begin{center}\rule{0.5\linewidth}{0.5pt}\end{center}

\hypertarget{declarations}{%
\subsection{Declarations}\label{declarations}}

\textbf{Funding.} This research received no external funding; it was
conducted as independent research at Tyche Institute.

\textbf{Competing interests.} The author declares no competing
interests. This work evaluates the open-source Project Veraison; the
author has no affiliation with, and claims no endorsement from, that
project.

\textbf{Generative-AI assistance.} The author used Anthropic's Claude
for drafting assistance, literature triage, and editorial review; all
claims, results, design decisions, and references are the author's own
and verified by the author, who takes full responsibility. The artifacts
and decoded results are released so that every reported value can be
independently reproduced. Reported per COPE and ICMJE recommendations.

\textbf{Author contributions (CRediT).} Anton Sokolov:
conceptualization, methodology, software, validation, investigation,
writing --- original draft, writing --- review and editing.

\textbf{Data and code availability.} See Section 7.1.

\hypertarget{references}{%
\subsection{References}\label{references}}

{[}1{]} H. Birkholz, D. Thaler, M. Richardson, N. Smith, and W. Pan,
\emph{Remote ATtestation procedureS (RATS) Architecture}, RFC 9334,
IETF, January 2023.

{[}2{]} L. Lundblade, G. Mandyam, J. O'Donoghue, and C. Wallace,
\emph{The Entity Attestation Token (EAT)}, RFC 9711, IETF, 2025.

{[}3{]} E. Voit, H. Birkholz, T. Hardjono, T. Fossati, and V. Scarlata,
\emph{Attestation Results for Secure Interactions (AR4SI)},
Internet-Draft \texttt{draft-ietf-rats-ar4si}, IETF, work in progress.

{[}4{]} T. Fossati, E. Voit, S. Trofimov, and H. Birkholz, \emph{EAT
Attestation Results (EAR)}, Internet-Draft \texttt{draft-ietf-rats-ear},
IETF, work in progress.

{[}5{]} H. Birkholz, T. Fossati, Y. Deshpande, and others, \emph{Concise
Reference Integrity Manifest (CoRIM)}, Internet-Draft
\texttt{draft-ietf-rats-corim}, IETF, work in progress.

{[}6{]} H. Birkholz, N. Smith, T. Fossati, and H. Tschofenig,
\emph{Remote ATtestation procedureS (RATS) Conceptual Message Wrapper
(CMW)}, RFC 9999, DOI 10.17487/RFC9999, July 2026.

{[}7{]} Project Veraison contributors, \emph{Veraison: an attestation
Verifier reference implementation}, source repository,
https://github.com/veraison/services (accessed June 2026).

{[}8{]} A. Sokolov, \emph{Composing Application-Layer Action Evidence
with Remote Attestation Procedures}, Internet-Draft
\texttt{draft-sokolov-rats-aep-composition-02}, IETF, 28 June 2026
(individual submission, not WG-adopted; expires 30 December 2026).

{[}9{]} A. Sokolov, \emph{When the Witness Is Also the Suspect:
Hardware-Rooting the Evidence That AI Agents Leave Behind}, IEEE
Internet Computing, under review (submitted 25 June 2026).

{[}10{]} A. Sokolov, \emph{Evidence Instrumentation for AI-Governance
Review: Three Public-Source Practitioner Use Cases}, IEEE Transactions
on Technology \& Society, MS 2026-06-0114-OTW-TTS, submitted 13 June
2026, rejected upon initial review 18 July 2026. Underlying data:
Zenodo, CC-BY, DOI 10.5281/zenodo.20488643.

{[}11{]} A. Sokolov, \emph{Hardware-rooted attestation for AI-agent
evidence: composing IETF RATS with action evidence packages}, Zenodo
preprint, CC-BY, concept DOI 10.5281/zenodo.20818671 (deposited 24 June
2026; current version 3, 31 July 2026).

{[}12{]} A. Sokolov, \emph{Policy-Bound Agent Evidence Packages} (AEP
attestation lab preprint v0.4), 27 May 2026.

{[}13{]} Trusted Computing Group, \emph{TPM 2.0 Library Specification},
TCG, latest revision.

{[}14{]} EnactTrust contributors, \emph{EnactTrust TPM attestation
scheme for Veraison (\texttt{tpm-enacttrust})}, profile
\texttt{https://enacttrust.com/veraison/1.0.0}. {[}15{]} R. Sailer, X.
Zhang, T. Jaeger, and L. van Doorn, ``Design and Implementation of a
TCG-based Integrity Measurement Architecture,'' in \emph{Proc. 13th
USENIX Security Symposium}, 2004, pp.~223--238.

{[}16{]} N. Schear, P. T. Cable II, T. M. Moyer, B. Richard, and R.
Rudd, ``Bootstrapping and Maintaining Trust in the Cloud,'' in
\emph{Proc. 32nd Annual Computer Security Applications Conference
(ACSAC)}, 2016, pp.~65--77. doi:10.1145/2991079.2991104.

{[}17{]} G. Coker, J. Guttman, P. Loscocco, A. Herzog, J. Millen, B.
O'Hanlon, J. Ramsdell, A. Segall, J. Sheehy, and B. Sniffen,
``Principles of remote attestation,'' \emph{International Journal of
Information Security}, vol.~10, no. 2, pp.~63--81, 2011.
doi:10.1007/s10207-011-0124-7.

{[}18{]} M. U. Sardar, A. Niemi, H. Tschofenig, and T. Fossati,
``Towards Validation of TLS 1.3 Formal Model and Vulnerabilities in
Intel's RA-TLS Protocol,'' \emph{IEEE Access}, vol.~12,
pp.~173670--173685, 2024. doi:10.1109/ACCESS.2024.3497184.

{[}19{]} M. U. Sardar, D. L. Quoc, and C. Fetzer, ``Towards
Formalization of Enhanced Privacy ID (EPID)-based Remote Attestation in
Intel SGX,'' in \emph{Proc. 23rd Euromicro Conference on Digital System
Design (DSD)}, 2020, pp.~604--607. doi:10.1109/DSD51259.2020.00099.

{[}20{]} M. Ambrosin, M. Conti, R. Lazzeretti, M. M. Rabbani, and S.
Ranise, ``Collective Remote Attestation at the Internet of Things Scale:
State-of-the-Art and Future Challenges,'' \emph{IEEE Communications
Surveys \& Tutorials}, vol.~22, no. 4, pp.~2447--2461, 2020.
doi:10.1109/COMST.2020.3008879.

{[}21{]} F. Mo, Z. Tarkhani, and H. Haddadi, ``Machine Learning with
Confidential Computing: A Systematization of Knowledge,'' \emph{ACM
Computing Surveys}, vol.~56, no. 11, art. 281, 2024.
doi:10.1145/3670007.

{[}22{]} A. Chan, C. Ezell, M. Kaufmann, K. Wei, L. Hammond, H. Bradley,
E. Bluemke, N. Rajkumar, D. Krueger, N. Kolt, L. Heim, and M.
Anderljung, ``Visibility into AI Agents,'' in \emph{Proc. 2024 ACM
Conference on Fairness, Accountability, and Transparency (FAccT)}, 2024,
pp.~958--973. doi:10.1145/3630106.3658948.

{[}23{]} A. Chan, K. Wei, S. Huang, N. Rajkumar, E. Perrier, S. Lazar,
G. K. Hadfield, and M. Anderljung, ``Infrastructure for AI Agents,''
\emph{Transactions on Machine Learning Research (TMLR)}, 2025.
https://openreview.net/forum?id=Ckh17xN2R2.

{[}24{]} Y. B. Lee, \emph{Permit Receipts for Permit-Before-Commit
Authorization of AI-Agent and Workload External Effects}, Internet-Draft
\texttt{draft-lee-orprg-permit-receipts-00}, IETF, 4 June 2026.

{[}25{]} M. U. Sardar, \emph{Pre-, Intra- and Post-handshake
Attestation}, Internet-Draft \texttt{draft-usama-seat-intra-vs-post-03},
IETF, 22 January 2026.

{[}26{]} T. Sato, \emph{Kernel Identity and Attestation for Governing
Enforcement Components}, Internet-Draft \texttt{draft-sato-soos-kia-02},
IETF, 10 June 2026.

{[}27{]} T. Sato, \emph{The Governance Audit Record (GAR) for Agentic AI
Systems}, Internet-Draft \texttt{draft-sato-soos-gar-02}, IETF, 10 June
2026.

{[}28{]} H. Birkholz et al., \emph{Verifiable Agent Conversations},
Internet-Draft \texttt{draft-birkholz-verifiable-agent-conversations},
IETF, work in progress.

{[}29{]} X. Carpent, N. Rattanavipanon, and G. Tsudik, ``Temporal
Consistency of Integrity-Ensuring Computations and Applications to
Embedded Systems Security,'' in \emph{Proc. ACM Asia Conf. on Computer
and Communications Security (ASIACCS)}, 2018. DOI:
10.1145/3196494.3196526.

{[}30{]} I. De Oliveira Nunes, S. Jakkamsetti, N. Rattanavipanon, and G.
Tsudik, ``On the TOCTOU Problem in Remote Attestation,'' in \emph{Proc.
ACM SIGSAC Conf. on Computer and Communications Security (CCS)}, 2021.
DOI: 10.1145/3460120.3484532.

\end{document}